\documentclass[aps, pra, twocolumn, reprint, noeprint, superscriptaddress,
]{revtex4-2}
\usepackage{times}
\usepackage{CJK} 
\usepackage{amsmath}
\usepackage{amssymb}
\usepackage{graphicx}
\usepackage{bm}
\usepackage{color}
\usepackage{subfigure}
\usepackage{booktabs} 
\usepackage{ulem}

\usepackage{mathrsfs}
\usepackage[colorlinks, 
            linkcolor=blue, 
            anchorcolor=blue, 
            citecolor=blue,
            ]{hyperref}

\begin{document}
\title{\texorpdfstring{Relativistic Calculations of Energy Levels, Field Shift Factors, and Polarizabilities \\ of Mercury and Copernicium}{Relativistic Calculations of Energy Levels, Field Shift Factors, and Polarizabilities of Mercury and Copernicium}}

\author{Hongxu Liu}
\thanks{These authors contributed equally to this work.}
\affiliation{State Key Laboratory of Metastable Materials Science and Technology
 \& Hebei Key Laboratory of Microstructural Material Physics,
 School of Science, Yanshan University, Qinhuangdao 066004, China}
\affiliation{Beijing National Laboratory for Condensed Matter Physics, Institute of Physics, Chinese Academy of Sciences, Beijing 100190, China}

\author{Jize Han}
\thanks{These authors contributed equally to this work.}
\affiliation{China Mobile (Suzhou) Software Technology Co., Ltd., Suzhou 215163, China}

\author{Yanmei Yu}
\email[Contact author: ]{ymyu@aphy.iphy.ac.cn}
\affiliation{Beijing National Laboratory for Condensed Matter Physics, Institute of Physics, Chinese Academy of Sciences, Beijing 100190, China}
\affiliation{University of Chinese Academy of Sciences, Beijing 100049, China}

\author{Yanfeng Ge}     
\affiliation{State Key Laboratory of Metastable Materials Science and Technology
 \& Hebei Key Laboratory of Microstructural Material Physics,
 School of Science, Yanshan University, Qinhuangdao 066004, China}

\author{Yong Liu}     
\affiliation{State Key Laboratory of Metastable Materials Science and Technology
 \& Hebei Key Laboratory of Microstructural Material Physics,
 School of Science, Yanshan University, Qinhuangdao 066004, China}

\author{Zhiguo Huang}
\affiliation{China Mobile (Suzhou) Software Technology Co., Ltd., Suzhou 215163, China}

\date{\today}
\begin{abstract}
Mercury (Hg) and superheavy element copernicium (Cn) are investigated using equation-of-motion relativistic coupled-cluster (EOM-RCC) and configuration interaction plus many-body perturbation theory (CI+MBPT) methods. Key atomic properties including ionization potentials (IP),  excitation energies (EEs), isotope field shift factors ($F$), and static electric dipole polarizabilities ($\alpha$) are calculated for ground and low-lying excited states. To evaluate the theoretical accuracy, calculations for both Hg and Cn are performed, with experimental data of Hg serving as benchmarks. Furthermore, basis set dependence has been systematically evaluated in the EOM-RCC calculations, with corresponding uncertainty estimates having been provided.
The calculated atomic properties could provide valuable insights into the electronic structure and chemical behavior of superheavy elements.
\end{abstract}

\maketitle

\section{Introduction}

Mercury (Hg) and superheavy element copernicium (Cn) are the two heaviest elements in Group~12 of the periodic table, and both play important roles in precision measurement. Hg, as the heaviest non-radioactive atom that can be laser-cooled and trapped, has been widely used in high-precision experiments, including parity violation tests and electric dipole moment (EDM) measurements. Its large nuclear charge enhances the sensitivity to possible physics beyond the Standard Model~\cite{lamoreaux1987new, romalis2001new, griffith2009improved, latha2009probing, loftus2011measurement, kumar2013working, bishof2016improved, graner2016reduced, sahoo2018relativistic}. Recent theoretical studies suggest that Cn exhibits an enhancement in the sensitivity factor of EDM by nearly an order of magnitude compared to Hg, making it a promising candidate for future EDM experiments~\cite{radvziute2016electric}. Hg is also considered a promising candidate for optical lattice quantum frequency standards. It offers projected systematic uncertainties below $10^{-18}$ and has a reduced blackbody radiation (BBR) shift coefficient relative to Sr and Yb systems \cite{hachisu2008trapping}. These characteristics make it suitable for applications such as detecting potential variations in the fine-structure constant over time~\cite{yamanaka2015frequency, takamoto2022perspective}.


Isotope shift (IS) studies of Hg and Cn also provide valuable input for testing and refining nuclear structure models. In Hg, high-precision IS measurements along the neutron-deficient isotopic chain reveal a shape staggering effect. Odd-mass isotopes such as $^{181}$Hg and $^{183}$Hg show abrupt increases in nuclear charge radii ($R_{\mathrm{ch}}$) relative to their even-mass neighbors~\cite{julin2016beam}. This pattern is interpreted as a quantum phase transition driven by the interplay between monopole and quadrupole nucleon interactions. The behavior is used to test the Monte Carlo nuclear shell model calculations~\cite{otsuka2001monte, borge2018focus}. In the superheavy region, the element Cn, first synthesized in 1996 at GSI Darmstadt~\cite{hofmann1996new}, is of particular interest. Its proximity to the predicted proton shell closure at $Z = 114$ makes it relevant to discussions of the next magic number in the nuclear shell model~\cite{sobiczewski1966closed, bender1999shell, sorlin2008nuclear}. Isotopes from $^{277}$Cn to $^{285}$Cn have been successfully produced~\cite{armbruster2011region, sobiczewski2011synthesis, munzenberg2015bohrium, saamark2021spectroscopy, utyonkov2015experiments}. The evolution of $R_{\mathrm{ch}}$ along the long Cn isotopic chain may reveal subtle effects associated with neutron filling beyond $N = 162$. These results can contribute to the understanding of nuclear structure near the region predicted as the ``island of stability'' around $Z = 114$ and $N = 184$~\cite{oganessian2004heavy, hamilton2013search}.



The ionization potentials (IP) and excitation energies (EEs) of Cn have been investigated in multiple studies~\cite{eliav1995transition, nash2005atomic, li2007atomic, dinh2008calculation, hangele2012accurate, lackenby2020calculation}, yet significant discrepancies exist among the reported values, which may complicate its spectroscopic identification. Beyond energy-related properties, other critical atomic characteristics of Cn remain insufficiently studied. Theoretical calculations of the electric dipole polarizability ($\alpha$), a key atomic parameter, have so far been restricted to the ground state~\cite{seth1997chemistry, roos2005new, pershina2008prediction, dzuba2016ionization, kumar2021relativistic, cheng2025relativistic}, with no reported data for the excited states. Additionally, theoretical predictions of the isotope field
shift factors ($F$), essential for atomic spectral analysis and $R_{\mathrm{ch}}$ extraction, are lacking. The absence of reliable data for these fundamental atomic properties, combined with inconsistencies in existing results, hinders the significance of a comprehensive investigation of this superheavy element. Systematic and precise theoretical calculations are thus necessary to fully characterize the atomic properties of Cn and provide reliable guidance for its experimental spectroscopic identification.

In this work, we employ two high-precision relativistic atomic calculation methods (EOM-RCC and CI+MBPT) to systematically investigate the electronic structure properties of group 12 elements Hg and its superheavy homologue Cn. This research focuses on key atomic properties of ground states and low-lying excited states in Hg and Cn, including IP, EEs, $F$ and $\alpha$. Initially, the reliability of the computational methods is verified through systematic comparisons between theoretical results and available experimental data for Hg. Building upon this validation, we extend the methodology to the superheavy region Cn, successfully obtaining key atomic parameters such as $F$ and $\alpha$ for low-lying excited states, which had not been previously reported in the literature to the best of our knowledge. Our calculations provide valuable theoretical references for future high-precision measurement on superheavy element Cn.

\section{Methods and computation strategies}

\subsection{The EOM-RCC Method}

The EOM-RCC calculation is based on the DIRAC code \cite{DIRAC24, shee2018equation}. The Dirac-Coulomb-Gaunt (DCG) Hamiltonian, expressed in atomic units, is given by,
\begin{eqnarray}
	\hat{H} &=& \sum_i \left[ c (\vec{\bm{\alpha}} \cdot \vec{\bf{p}})_i + (\bm{\beta} - 1)_i m_0 c^2 + V_{\text{nuc}}(r_i) \right] \nonumber \\
	&& + \sum_{i<j} \left[ \frac{1}{r_{ij}} - \frac{1}{2} \frac{\vec{\bm{\alpha}}_i \cdot \vec{\bm{\alpha}}_j}{r_{ij}} \right], \label{DCG}
\end{eqnarray}
where $\vec{\bm{\alpha}}$ and $\bm{\beta}$ are the Dirac matrices, $\vec{\mathbf{p}}$ denotes the momentum operator, $m_0 c^2$ is the electron rest mass energy, and $c$ is the speed of light. The term $V_{\text{nuc}}(r_i)$ represents the nuclear potential, while $r_{ij}$ is the distance between the $i$-th and $j$-th electrons. The final term includes the Gaunt interaction, which is the dominant component of the Breit interaction.

Progress has been made in the exact two-component (X2C) method in recent years, which reduces the four-component Dirac equation to a two-component form, preserving key relativistic effects while enhancing computational efficiency~\cite{liu2020essentials}. Recently, Knecht~\textit{et al.} extended the atomic mean-field model to incorporate both scalar-relativistic and spin--orbit two-electron picture-change effects (PCEs) within the X2C Hamiltonian formalism. This model, constructed using the DCG Hamiltonian, is referred to as eamfX2C$_{\rm DC}$~\cite{knecht2022exact}. In this work, the eamfX2C$_{\rm DC}$ Hamiltonian is employed to calculation the IP, EEs and $F$ factors for the Hg and Cn atoms. For the finite-field calculation of $\alpha$, an earlier version of the atomic mean-field X2C Hamiltonian is used. This version, AMFI X2C$_{\rm D}$, applies a molecular mean-field approach to the two-electron terms~\cite{sikkema2009molecular}. The effective quantum electrodynamics (effQED) potential implemented by Sunaga~\textit{et al.}~\cite{sunaga20224} is adopted. It includes the Uehling potential for vacuum polarization~\cite{uehling1935polarization}, the self-energy model potential proposed by Pyykkö and Zhao~\cite{pyykko2003search}, and the effective self-energy potential developed by Flambaum and Ginges (FG)~\cite{flambaum2005radiative}. To treat QED effects in a more rigorous manner, the effQED potentials are incorporated into the four-component DCG calculations. Both the Uehling vacuum polarization and FG self-energy potentials are added to the DCG Hamiltonian and treated self-consistently.

The EOM-RCC method is based on a Dirac--Hartree--Fock (DHF) reference state $\left|\Phi_0\right\rangle$ corresponding to the closed-shell ground state of Hg and Cn. The ground-state correlation is first described using an exponentiated coupled-cluster operator $\hat{T}$, which yields the correlated ground state $\left|\Psi_0\right\rangle = e^{\hat{T}} \left|\Phi_0\right\rangle$. An excited state $\left|\Psi_k\right\rangle$ is then obtained by applying a linear excitation operator $\hat{R}_k$ to the correlated ground state, 
\[
\left|\Psi_k\right\rangle = \hat{R}_k e^{\hat{T}} \left|\Phi_0\right\rangle.
\]
The EEs and the amplitudes of the $\hat{R}_k$ operator are obtained by solving the equation of motion,
\begin{equation}
\bar{H} \hat{R}_k \left|\Phi_0\right\rangle = \omega_k \hat{R}_k \left|\Phi_0\right\rangle,
\end{equation}
where $\bar{H} = e^{-\hat{T}} H e^{\hat{T}}$ is the similarity-transformed Hamiltonian, and $\omega_k$ is the excitation energy. The EOM-RCC method considers single and double excitations. For Hg ($Z = 80$), we consider two electron spaces: the $4d^{10}5s^2 5p^6 5d^{10}6s^2$ (e44) and $4s^2 4p^6 4d^{10}5s^2 5p^6 5d^{10}6s^2$ (e52). We find that including additional inner-shell electrons, such as in the $3d^{10}4s^2 4p^6 4d^{10}5s^2 5p^6 5d^{10}6s^2$ (e62) configuration, does not affect the results at the current level of accuracy, as shown in Table~\ref{tab:Hg-core} of Appendix. Therefore, all final EOM-RCC calculations for Hg are carried out using the e44 active space. For Cn, we adopt a similar strategy and use the $5d^{10}5f^{14}6s^2 6p^6 6d^{10}7s^2$ (e44) active space for the EOM-RCC calculations.

The EOM-RCC calculations are performed using the Dyall family of basis sets, which are developed for heavy $s$- and $d$-block elements. These basis sets are available in valence (vNz), core-valence (cvNz), and all-electron (aeNz) forms. The cardinal number $N = 2$, $3$ or $4$ corresponds to double-zeta ($2\zeta$), triple-zeta ($3\zeta$) and quadruple-zeta ($4\zeta$) qualities, respectively~\cite{dyall2010revised,dyall2011relativistic}. To improve the description of diffuse electron density, we adopt the recently developed augmented Dyall basis sets. These include newly optimized diffuse $s$, $p$, and $d$ functions, along with the previously available diffuse $f$ and $g$ functions. The resulting augmented all-electron basis sets, denoted as dyall.aaeNz, provide improved representation of the outer electron cloud~\cite{dyall2022diffuse}. This is relevant for the accurate calculation of properties such as electron affinities, polarizabilities and EEs. Benchmark calculations indicate that these standard augmented basis sets (dyall.aaeNz) do not fully capture the behavior of highly excited states, such as the $6s7s$ states in Hg. Deviations from experimental EEs are observed, particularly for states with diffuse character. To address this, we construct extended basis sets by adding an additional layer of diffuse functions for each angular momentum type ($s$, $p$, $d$, $f$, and $g$), generated using an even-tempered formula. The resulting singly-augmented (s-aug) basis sets are referred to as s-aug-dyall.aaeNz. The use of these additional diffuse functions improves the computed EEs for highly excited states and leads to closer agreement with experimental values. Given the similarity in electronic structure, the same augmentation approach is applied to Cn.

To evaluate convergence of EEs with respect to diffuse functions, doubly-augmented (d-aug) basis sets are employed, which contain two sets of diffuse functions for each angular momentum. This allows evaluation of basis set dependence for properties sensitive to diffuse electron density. All calculations are carried out using the DIRAC program package. The DCG Hamiltonian is employed with uncontracted basis sets. The small-component basis functions are generated using the restricted kinetic balance condition. Electron correlation is treated at the EOM-RCC level. The full positive-energy virtual space is included in the correlation treatment, except for virtual orbitals with energies above 100~a.u.. Convergence tests indicate that this energy cutoff can be neglected at the current level of accuracy in the calculated properties. To calculate the scalar and tensor polarizabilities ($\alpha^S$ and $\alpha^T$), a finite-field perturbation is introduced into the Hamiltonian through the term $\vec{D} \cdot \vec{\mathcal{E}}$, where $\vec{D}$ is the electric dipole operator and $\vec{\mathcal{E}}$ is the applied electric field. Total energies are calculated for a range of field strengths $|\vec{\mathcal{E}}|$. These data are fitted to a Taylor expansion centered at $|\vec{\mathcal{E}}| = 0$. The second-order term of the expansion yields the polarizability $\alpha(J_n, M_{J_n})$ for each magnetic sublevel $M_{J_n}$. From these values, the $\alpha^S$ and $\alpha^T$ components are extracted.

In the case of $F$ factors, one-electron operators for Fermi-contact integrals corresponding to a specific nucleus are added to the DHF Hamiltonian. The electron density at the nucleus ($\rho_0$) is then obtained by fitting the finite-field energies. The $F$ factor is evaluated using,
\begin{equation}
F = \frac{Z e^2 \Delta \rho_0}{6 \epsilon_0},
\end{equation}
where $Z$ is the nuclear charge number, $e$ is the elementary charge, $\epsilon_0$ is the vacuum permittivity, and $\Delta \rho_0$ represents the change in electron density at the nucleus between two states. After verifying the convergence of our results with respect to the freezing of inner-core electrons and the truncation of the virtual orbital space, we identify the basis set incompleteness as the dominant source of uncertainty in the EOM-RCC calculations. Results from three basis sets (s-aug-dyall.aae3z, s-aug-dyall.aae4z, and d-aug-dyall.aae4z) are compared of quantify basis set errors. The uncertainty is defined as the sum of absolute differences between adjacent basis sets: $\lvert 4\zeta^{s}-3\zeta^{s} \rvert+\lvert4\zeta^{d}-4\zeta^{s}\rvert$. This approach accounts for both basis set size and diffuse function effects on the calculation accuracy.

\begin{table*}[htbp]
	\caption{Energy levels (in cm$^{-1}$) of Hg are calculated using the EOM-RCC and CI+MBPT methods and are compared with experimental values~\cite{kramida2020nist}.
 \label{tab:HgEE}}
	{\setlength{\tabcolsep}{7pt}
		\begin{tabular}{llllllllllllllllllll}\hline\hline\addlinespace[0.2cm]
	&		&	 \multicolumn{7}{c}{EOM-RCC}													&	&	 \multicolumn{3}{c}{CI+MBPT}					&		\\  \cline{3-9} \cline{11-13}  \addlinespace[0.2cm]
Conf.	&	$LS$	&	$2\zeta^{s}$	&	$3\zeta^{s}$	&	$4\zeta^{s}$	&	$4\zeta^{d}$	&	$\Delta_{\rm QED}$	&	Final	&	Unc. 	&	&	$\Delta_{\rm QED}$	&	Final	&	Unc. 	&	NIST \cite{kramida2020nist} 	\\ \hline\addlinespace[0.2cm]
IP																										\\
$6s^2$	&	$^1S_0$	&	83239 	&	83813 	&	84066 	&	84161 	&	-176 	&	83985 	&	348 	&	&		&		&		&	84184.15	\\
EE																										\\
$6s6p$	&	$^3P_0$	&	37516 	&	37223 	&	37454 	&	37498 	&	-187 	&	37312 	&	275 	&	&	-194 	&	37657 	&	1754 	&	37644.982	\\
	&	$^3P_1$	&	39235 	&	39009 	&	39249 	&	39298 	&	-184 	&	39114 	&	289 	&	&	-192 	&	39531 	&	1689 	&	39412.237	\\
	&	$^3P_2$	&	43458 	&	43492 	&	43777 	&	43843 	&	-175 	&	43669 	&	351 	&	&	-186 	&	44548 	&	1502 	&	44042.909	\\
	&	$^1P_1$	&	54541 	&	54445 	&	54712 	&	54779 	&	-164 	&	54615 	&	334 	&	&	-174 	&	54736 	&	1527 	&	54068.6829	\\
$6s7s$	&	$^3S_1$	&	61464 	&	61886 	&	62217 	&	62283 	&	-158 	&	62125 	&	396 	&	&	-173 	&	66488 	&	931 	&	62350.325	\\
	&	$^1S_0$	&	63358 	&	63812 	&	64152 	&	64220 	&	-156 	&	64064 	&	408 	&	&	-173 	&	68311 	&	1048 	&	63928.12	\\
$5d^9 6s^2 6p$	&	$^3P_2$	&	67446 	&	67582 	&	68302 	&	68444 	&	35 	&	68479 	&	863 	&	&	-15 	&	73326 	&	4651 	&	68886.43	\\
$6s7p$	&	$^3P_0$	&	68944 	&	69660 	&	69741 	&	69804 	&	-172 	&	69632 	&	144 	&	&	-189 	&	73691 	&	989 	&	69516.576	\\
	&	$^3P_1$	&	69059 	&	69785 	&	69883 	&	69948 	&	-170 	&	69778 	&	164 	&	&	-188 	&	73839 	&	1022 	&	69661.803	\\
$5d^9 6s^2 6p$	&	$^3D_3$	&	69542 	&	69425 	&	70266 	&	70422 	&	100 	&	70522 	&	996 	&	&	85 	&	75454 	&	5506 	&	70932.2	\\ \hline\hline
	\end{tabular}}
\end{table*}

\begin{table*}[htbp]
	\caption{Energy levels (in cm$^{-1}$) of Cn are calculated using the EOM-RCC and CI+MBPT methods. The results are compared with previously reported theoretical values. 
 \label{tab:CnEE}}
	{\setlength{\tabcolsep}{3.4pt}
		\begin{tabular}{lllllllllllllllllllllllll}\hline\hline\addlinespace[0.2cm]
	&		&	 \multicolumn{7}{c}{EOM-RCC}													&	&	 \multicolumn{3}{c}{CI+MBPT}					&	&	 \multicolumn{4}{c}{Refs}							\\  \cline{3-9} \cline{11-13} \cline{15-18} 
Conf.	&	$LS$	&	$2\zeta^{s}$	&	$3\zeta^{s}$	&	$4\zeta^{s}$	&	$4\zeta^{d}$	&	$\Delta_{\rm QED}$	&	Final	&	Unc.	&	&	$\Delta_{\rm QED}$	&	Final	&	Unc.	&	&	\cite{li2007atomic}	&	\cite{dinh2008calculation}	&	\cite{hangele2012accurate}	&		\\ \hline
IP																																	\\ 
$6d^{10} 7s^2$	&	$^1S_0$	&	94341 	&	95501 	&	96107 	&	96151 	&	131 	&	96238 	&	650 	&	&		&		&		&	&		&		&	91569 	&	96545 \cite{eliav1995transition}	\\ 
	&		&		&		&		&		&		&		&		&	&		&		&		&	&		&		&		&	97956 \cite{lackenby2020calculation}	\\ 
EE																																	\\ 
$6d^9 7s^2 6p$	&	$^3P_2$	&	34530 	&	35011 	&	35590 	&	35584 	&	137 	&	35721 	&	585 	&	&	141 	&	34482 	&	2016 	&	&	34150 	&	35785 	&		&		\\
	&	$^3F_3$	&	37682 	&	38086 	&	38569 	&	38540 	&	140 	&	38680 	&	511 	&	&	144 	&	37502 	&	1882 	&	&	37642 	&	38652 	&		&		\\
$6d^{10} 7s7p$	&	$^3P_0$	&	50006 	&	49853 	&	50172 	&	50196 	&	-360 	&	49835 	&	342 	&	&	-386 	&	52037 	&	2845 	&	&	48471 	&	51153 	&	40792 	&		\\
	&	$^3P_1$	&	51961 	&	52012 	&	52415 	&	52441 	&	-244 	&	52198 	&	430 	&	&	-311 	&	53943 	&	1904 	&	&	52024 	&	55057 	&	45705 	&		\\
$6d^9 7s^2 7p$	&	$^3F_4$	&	57355 	&	57997 	&	58728 	&	58777 	&	141 	&	58918 	&	781 	&	&	143 	&	58308 	&	1839 	&	&	60366 	&	56131 	&		&		\\
	&	$^3D_2$	&	57854 	&	58477 	&	59175 	&	59210 	&	140 	&	59349 	&	733 	&	&	141 	&	58829 	&	1703 	&	&	60809 	&	56960 	&		&		\\
	&	$^3P_1$	&	59645 	&	60139 	&	60826 	&	60868 	&	73 	&	60941 	&	729 	&	&	104 	&	60857 	&	1294 	&	&	64470 	&	58260 	&		&		\\
	&	$J=3$	&	60425 	&	61045 	&	61789 	&	61838 	&	140 	&	61978 	&	793 	&	&	142 	&	61367 	&	1797 	&	&		&		&		&		\\
	&	 $J=2$	&	61249 	&	61859 	&	62434 	&	62414 	&	205 	&	62619 	&	596 	&	&	219 	&	62032 	&	1345 	&	&		&		&		&		\\ 
	&	$^3P_1$	&	67764 	&	68221 	&	68749 	&	68724 	&	96 	&	68820 	&	553 	&	&	107 	&	68588 	&	1372 	&	&	73686 	&	68673 	&		&		\\
$6d^{10} 7s8s$	&	$^1S_0$	&		&	71772 	&	72643 	&	72737 	&	173 	&	72910 	&	966 	&	&	150 	&		&	2096 	&	&		&	88861 	&	75845 	&		\\
	&	$^3S_1$	&	71018 	&	72056 	&	72925 	&	73019 	&	174 	&	73193 	&	963 	&	&	149 	&		&	1817 	&	&		&	87785 	&	71964 	&		\\
$6d^{10} 7s7p$	&	$^3P_2$	&	73356 	&	73332 	&	73794 	&	73876 	&	-367 	&	73509 	&	544 	&	&	-391 	&	76526 	&	2420 	&	&	76641 	&	73736 	&	58590 	&		\\
	&	$^1P_1$	&	79314 	&	79624 	&	80070 	&	80158 	&	-271 	&	79887 	&	534 	&	&	-271 	&	84093 	&	981 	&	&	85533 	&	79637 	&	66089 	&		\\\hline\hline
    \end{tabular}}
\end{table*}

\subsection{The CI+MBPT Method}

The CI+MBPT method employed in this work is recognized as an advanced and versatile approach that rigorously accounts for both valence electrons and holes in nearly filled shells. This particle--hole formalism is particularly suitable for heavy and superheavy elements, where excited configurations such as $5d^9$ in Hg and $6d^9$ in Cn can be described as hole states within the fully occupied $5d$ and $6d$ subshells, respectively. The CI+MBPT framework supports the treatment of systems with an arbitrary number of valence electrons and holes. It has been applied to a wide range of heavy atomic systems~\cite{kahl2021ab}.

A $V^{N-2}$ potential is adopted to perform the DHF calculations for constructing the single-electron orbital basis. In this approximation, the closed-shell core is defined by removing the outermost $ns^2$ electrons ($n = 6$ for Hg and $n = 7$ for Cn). The DHF calculation includes both the Uehling vacuum polarization potential~\cite{ginges2016qed} and the finite-grid self-energy potential~\cite{ginges2016atomic}. The $V^{N-2}$ approximation is selected for its simplicity in including core--valence correlations, which are subsequently treated using MBPT. The single-particle orbital basis is generated by diagonalizing a set of $B$ splines over the DHF potential. Configurations are included by allowing all SD excitations from the following reference configuration sets,
\begin{itemize}
    \item Hg: $6s^2$, $6s^1 6p^1$ and $5d^{-1}6s^2 6p$
    \item Cn: $7s^2$, $7s^1 7p^1$ and $6d^{-1}7s^2 7p^1$.
\end{itemize}
The excitations are restricted to $n$--$spdf$ orbitals, with principal quantum numbers $n < 12$, $18$ and $21$, respectively. To reduce the size of the configuration interaction (CI) matrix and the computational cost, the emu CI method~\cite{kahl2019ambit, geddes2018saturated} is used. This method is based on the observation that the wave functions of interest are primarily determined by a subset $N_{\mathrm{small}}$ of low-lying configurations. These configurations define the effective CI matrix~\cite{dzuba2017combining}. Off-diagonal matrix elements that do not involve at least one of the dominant configurations are set to be zero. These elements have negligible effect on the eigenstates of interest and without any loss of accuracy. 

For both Hg and Cn, the dominant configurations are limited to those generated by single excitations up to $nspdf$, and by single and double excitations up to $7spdf$ for Hg and $8spdf$ for Cn. As the principal quantum number $n$ increases through $12$, $18$, and $21$, the EEs remain nearly unchanged. In contrast, the inclusion of high-angular-momentum orbitals up to $12spdfg$ leads to noticeable changes in the EEs. This variation is treated as the uncertainty associated with the CI+MBPT results. Core--valence correlations are included through MBPT~\cite{dzuba1996combination, berengut2006calculation}, incorporating all one-, two-, and three-body diagrams. Orbitals up to $35spdfgh$ ($n \leq 35$, $0 \leq l \leq 5$) are included. Subtraction diagrams in MBPT are also considered. The use of the $V^{N-2}$ potential, instead of the $V^N$ potential, is supported by better agreement with experimental EEs in the case of Hg. To determine the $F$ factors, the nuclear root-mean-square charge radius $r_c$ is varied. Specifically, values of $r_c = 5.8$, $6.3$, and $6.8$~fm are used for Hg, and $r_c = 6.358$, $6.858$ and $7.358$~fm are used for Cn. The $F$ factors are then fitted to the CI+MBPT results using the linear relation
\begin{equation}
v^{\rm FIS} = F\, \delta\langle r^2 \rangle,
\end{equation}
as described in Ref.~\cite{berengut2003isotope}.

\section{Results and Discussion}

\begin{table}[htbp]
	\caption{The isotope field shift ($F$) factors (in GHz/fm$^2$) for transitions between the ground state and the ten low-lying excited states in Hg and Cn, calculated using the EOM-RCC and CI+MBPT methods. \label{tab:fs}}
	{\setlength{\tabcolsep}{3pt}
		\begin{tabular}{llllllllllllllllllll}\hline\hline\addlinespace[0.2cm]
												
Conf.	&	$LS$	&	 \multicolumn{3}{c}{EOM-RCC}					&	&	CI+MBPT 	\\\cline{3-5}
	&		&	$3\zeta^{s}$	&	$4\zeta^{s}$	&	$4\zeta^{d}$	&	&		\\\hline 
Hg	&		&		&		&		&	&		\\
$6s6p$	&	$^3P_0$	&	-55.26 	&	-55.62 	&	-55.76(50)	&	&	-55.72(86)	\\
	&	$^3P_1$	&	-55.47 	&	-55.86 	&	-56.01(54)	&	&	-55.97(86)	\\
	&	$^3P_2$	&	-55.27 	&	-55.86 	&	-56.03(76)	&	&	-56.10(86)	\\
	&	$^1P_1$	&	-49.61 	&	-50.14 	&	-50.28(67)	&	&	-50.65(78)	\\
$6s7s$	&	$^3S_1$	&	-45.93 	&	-46.39 	&	-46.52(59)	&	&	-46.40(76)	\\
	&	$^1S_0$	&	-45.30 	&	-45.76 	&	-45.89(59)	&	&	-45.88(76)	\\
$5d^9 6s^2 6p$	&	$^3P_2$	&	16.20 	&	11.07 	&	 10.3(5.9)	&	&	-8.66(36)	\\
$6s7p$	&	$^3P_0$	&	-50.22 	&	-50.72 	&	-50.86(64)	&	&	-50.92(79)	\\
	&	$^3P_1$	&	-49.76 	&	-50.33 	&	-50.49(73)	&	&	-51.54(11)	\\
$5d^9 6s^2 6p$	&	$^3D_3$	&	29.38 	&	29.57 	&	 29.46(31)	&	&	23.25(40)	\\ \hline
Cn	&		&		&		&		&	&		\\
$6d^9 7s^2 6p$	&	$^3P_2$	&	297.94 	&	298.36 	&	 298.56(62)	&	&	262.46(15)	\\
	&	$^3F_3$	&	314.13 	&	314.66 	&	 314.74(61)	&	&	270.77(15)	\\
$6d^{10} 7s7p$	&	$^3P_0$	&	-822.56 	&	-828.48 	&	-829.00(52)	&	&	-803.00(46)	\\
	&	$^3P_1$	&	-581.04 	&	-599.14 	&	-599.47(33)	&	&	-318.51(9)	\\
$6d^9 7s^2 7p$	&	$^3F_4$	&	232.93 	&	231.69 	&	 231.26(43)	&	&	187.11(11)	\\
	&	$^3D_2$	&	227.02 	&	227.84 	&	 228.82(98)	&	&	183.79(11)	\\
	&	$^3P_1$	&	95.13 	&	101.56 	&	 101.93(37)	&	&	-91.48(10)	\\
	&	$J=3$	&	243.37 	&	242.09 	&	 241.62(47)	&	&	193.27(71)	\\
	&	 $J=2$	&	349.52 	&	346.78 	&	 345.56(122)	&	&	300.71(17)	\\
	&	$^3P_1$	&	110.39 	&	112.65 	&	 112.99(34)	&	&	35.43(3)	\\
$6d^{10} 7s8s$	&	$^1S_0$	&	388.38 	&	387.88 	&	 389.44(56)	&	&		\\
	&	$^3S_1$	&	388.53 	&	388.03 	&	 387.68(85)	&	&		\\
$6d^{10} 7s7p$	&	$^3P_2$	&	-899.06 	&	-904.73 	&	-905.92(119)	&	&	-766.74(30)	\\
	&	$^1P_1$	&	-635.54 	&	-626.45 	&	-628.16(171)	&	&		\\\hline\hline

	\end{tabular}}
\end{table}

\begin{table}[htbp]
	\caption{The isotope field shift ($F$) factors (in GHz/fm$^2$) for the transitions $6s^2\,{}^1S_0 \rightarrow 6s6p\,{}^3P_0$ ($\lambda = 265$~nm), $6s^2\,{}^1S_0 \rightarrow 6s6p\,{}^3P_1$ ($\lambda = 254$~nm), and $6s6p\,{}^3P_2 \rightarrow 6s7s\,{}^3S_1$ transition ($\lambda = 546$~nm) in the Hg atom, calculated using the EOM-RCC and CI+MBPT methods. The results are compared with previously reported theoretical and experimental values. Abbreviations: MCDHF - Multiconfiguration Dirac–Hartree–Fock; MCDF - Multiconfiguration Dirac–Fock. \label{tab:fsHg}}
	{\setlength{\tabcolsep}{12pt}
		\begin{tabular}{lllllllllllll}\hline\hline\addlinespace[0.2cm]
        
Transition	&	$F$ factor	&	Method	\\\hline 
265 nm	&	-55.76(50)	&	EOM-RCC (This work)	\\
	&	-55.72(86)	&	CI+MBPT (This work)	\\
	&	-57.71(99)	&	MCDHF \cite{schelfhout2022multiconfiguration} 	\\ 
254 nm	&	-56.01(54)	&	EOM-RCC (This work)	\\
	&	-55.97(86)	&	CI+MBPT (This work)	\\
	&	-55.36	&	MCDF \cite{torbohm1985state}	\\
	&	-58.07(1.00) 	&	MCDHF \cite{schelfhout2022multiconfiguration} 	\\
	&	-64.8(6.0)	&	semi-emp \cite{ulm1986isotope}	\\
	&	-55.1(3.1)	&	electronic X-ray \cite{lee1978variations}	\\
	&	-52.1(1.0)	&	muonic X-ray \cite{hahn1979experimental}	\\
546 nm	&	9.51(96)	&	EOM-RCC (This work)	\\
	&	9.7(1.1)	&	CI+MBPT (This work)	\\
	&	9.73	&	semi-emp. \cite{ulm1986isotope}	\\\hline\hline

	\end{tabular}}
\end{table}

\begin{table}[htbp]
	\caption{The static scalar ($\alpha^S$) and tensor ($\alpha^T$) electric dipole polarizabilities (in a.u.) of the ground and excited states in Hg and Cn, calculated using the EOM-RCC method. The uncertainty is estimated from the difference between the $3\zeta^s$ and $4\zeta^s$ basis sets and is given in parentheses. Abbreviations: Expt. – Experiment; PRCCSD(T) - Perturbative Relativistic Coupled-Cluster with Single and Double excitations and perturbative Triples; RNCCSD - Relativistic Normal Coupled-Cluster with Single and Double excitations; RPA: Random Phase Approximation.
\label{tab:alpha}}
	{\setlength{\tabcolsep}{2.2pt}
		\begin{tabular}{llrlllllllll}\hline\hline\addlinespace[0.2cm]	
        
Conf.	&	$LS$	&		&	\multicolumn{2}{c}{EOM-RCC}			&	\multicolumn{1}{c}{Refs}		\\ \cline{4-5}
	&		&		&	$3\zeta^{s}$	&	$4\zeta^{s}$				\\\hline
Hg												\\
$6s^2$	&	$^1S_0$	&	$\alpha^S$	&	35.53 	&	35.28(25)	&	33.91(34),	Expt. \cite{goebel1996dipole}	\\
	&		&		&		&		&	33.75,	Expt. \cite{tang2008dynamical}	\\
	&		&		&		&		&	34.15,	CCSD(T) \cite{pershina2008prediction}	\\
	&		&		&		&		&	33.6,	CI+MBPT \cite{hachisu2008trapping}	\\
	&		&		&		&		&	34.27,	PRCCSD(T) \cite{singh2015rigorous}	\\
	&		&		&		&		&	34.1,	CCSD(T) \cite{borschevsky2015high}	\\
	&		&		&		&		&	34.2(5),	RNCCSD \cite{sahoo2018relativistic}	\\
	&		&		&		&		&	34.5(8),	PRCCSD(T) \cite{sahoo2018role}	\\
	&		&		&		&		&	33.69(34),	PRCCSD(T) \cite{kumar2021relativistic}	\\
	&		&		&		&		&	34.04(68),	CCSD(T) \cite{cheng2025relativistic}	\\
$6s6p$	&	$^3P_0$	&	$\alpha^S$	&	56.40 	&	56.03(37)	&	54.6,	CI+MBPT \cite{hachisu2008trapping}	\\
	&	$^3P_1$	&	$\alpha^S$	&	58.97 	&	58.60(37)	&			\\
	&		&	$\alpha^T$	&	9.63 	&	9.54(9)	&			\\
	&	$^3P_2$	&	$\alpha^S$	&	89.04 	&	88.79(25)	&			\\
	&		&	$\alpha^T$	&	-29.35 	&	-29.24(11)	&			\\
\multicolumn{2}{r}{$^3P_0-^1S_0$}			&	$\Delta\alpha^S$	&	20.87 	&	20.75(45)	&	21.0,	CI+MBPT \cite{hachisu2008trapping}	\\\hline
Cn												\\
$7s^2$	&	$^1S_0$	&	$\alpha^S$	&	28.46 	&	27.99(47)	&	25.82,	CCSD(T) \cite{seth1997chemistry}	\\
	&		&		&		&		&	28.68,	CCSD(T) \cite{roos2005new}	\\
	&		&		&		&		&	27.40,	CCSD(T) \cite{pershina2008prediction}	\\
	&		&		&		&		&	28(4),	RPA \cite{dzuba2016ionization}	\\
	&		&		&		&		&	27.44(88),	PRCCSD(T) \cite{kumar2021relativistic}	\\
	&		&		&		&		&	27.92(28),	CCSD(T) \cite{cheng2025relativistic}	\\
$6d^9 6p$	&	$^3P_2$	&	$\alpha^S$	&	35.44 	&	34.75(69)	&			\\
	&		&	$\alpha^T$	&	-0.76 	&	-0.77(1)	&			\\
	&	$^3F_3$	&	$\alpha^S$	&	37.39 	&	36.55(84)	&			\\
	&		&	$\alpha^T$	&	-0.02 	&	-0.30(28)	&			\\
$7s7p$	&	$^3P_0$	&	$\alpha^S$	&	36.89 	&	34.78(2.11)	&			\\
	&	$^3P_1$	&	$\alpha^S$	&	42.97 	&	38.99(3.02)	&			\\
	&		&	$\alpha^T$	&	1.53 	&	1.48(5)	&			\\
\multicolumn{2}{r}{$^3P_0-^1S_0$}			&	$\Delta\alpha^S$	&	8.43 	&	6.79(1.64)	&			\\\hline

\end{tabular}}
\end{table}

Table~\ref{tab:HgEE} presents the calculated energy levels of Hg using the EOM-RCC and CI+MBPT methods. In the EOM-RCC method, the energy of the ground state is expressed through the IP, defined as the energy difference between the total energy of Hg$^+$ and Hg computed with single and double excitations. Using the s-aug-dyall.aaeNz basis sets, the calculated IP and EEs show alignment with reference values, even when smaller basis sets such as $2\zeta^s$ are applied. As the basis set size increases to $3\zeta^s$ and $4\zeta^s$, a consistent trend of convergence is observed. The calculated data agree with experimental values~\cite{kramida2020nist} for both two-valence-electron states, such as $6sns$ and $6snp$; and $d^9$-hole valence states, such as $5d^9 6s^2 6p$. Replacing the s-aug-dyall.aae4z with the d-aug-dyall.aae4z basis sets leads to minor changes in the calculated IP and EE values. These changes are smaller than those introduced by increasing the basis size from $3\zeta^s$ to $4\zeta^s$. Therefore, instead of performing a basis set extrapolation, the results obtained with the $4\zeta^d$ basis and the QED corrections $\Delta_{\rm QED}$ are adopted as the final values. Uncertainties are estimated from the sum of differences: $\lvert 4\zeta^{s}-3\zeta^{s} \rvert+\lvert4\zeta^{d}-4\zeta^{s}\rvert$. While the completeness of the active coupled-cluster space has been checked by confirming convergence with respect to the treatment of core electrons and the truncation of virtual orbitals, the finite size of the basis set is identified as the dominant source of uncertainty. The relatively large deviation indicates a limitation in Dyall's basis, which ends with the 4$\zeta$ level. Adding diffuse functions improves the result but does not eliminate the basis size dependence. Results from the CI+MBPT method are consistent for lower excited states, including $6s6p~^3P_{0,1,2}$ and $^1P_1$. For higher excited states, such as $6s7s$, $6s7p$, and $5d^9 6s^2 6p$, the discrepancies between computed and reference values become more obvious. This behavior is attributed to the treatment of $d^{-1}$-hole states, a factor previously identified in related studies~\cite{dinh2008calculation}. The QED corrections $\Delta_{\rm QED}$ to the EEs generally follow expected trends. An exception is found in the $5d^9 6s^2 6p~^3P_2$ state, where the correction has a reversed sign and smaller magnitude compared to the other states.

Table~\ref{tab:CnEE} presents the calculated energy levels of Cn obtained using the same calculation methodology as applied to Hg. The results suggest that basis set size effects constitute the primary source of uncertainty, with a larger impact observed for Cn relative to Hg. This behavior implies that treatment of electron correlation in heavier atomic systems requires larger basis sets. For the IP, the present EOM-RCC calculations show agreement with reference values from the Fock-space coupled cluster (FSCC) results of Eliav et al.~\cite{eliav1995transition} and the configuration interaction perturbation theory (CIPT) calculations reported by Lackenby et al.~\cite{lackenby2020calculation}. Concerning the EEs, both EOM-RCC and CI+MBPT methods yield consistent results for low-lying states, with calculated values in accord with those reported by Li et al.~\cite{li2007atomic} and Dinh et al.~\cite{dinh2008calculation}. For higher excitation states, notably the $6d^{9}7s^27p$ configuration, larger deviations from reference values emerge. The EOM-RCC and CI+MBPT approaches maintain internal consistency. For the $7s8s$ Rydberg state, the EOM-RCC results show agreement with the experimental data of H\"{a}ngele et al.~\cite{hangele2012accurate}, whereas the CI+MBPT approach appears less effective in reproducing the experimental energies---a trend consistent with observations for Hg. The $\Delta_{\rm QED}$ from both methods display consistent behavior across all investigated electronic states.

\begin{table*}[htp]
\centering		
\caption{The excitation energies (EE) of the ground and lower excited states of Hg atoms are calculated using EOM-RCC and different correlation models (in cm$^{-1}$). We considered more inner shell electron correlation models under the dyall.ae2z-lt10 model, where lt10 means the virtual orbital cutoff energy is 10 a.u.. e34, which means the number of inner shell correlation electrons is 34 (including 4f5s5p5d6s orbital electrons). \label{tab:Hg-core}}{\setlength{\tabcolsep}{12pt}
\begin{tabular}{lllllll}\hline\hline  \addlinespace[0.1cm]
	&		&	 \multicolumn{4}{c}{dyall.ae2z-\texttt{lt10}}							&		\\  \cline{3-6}  \addlinespace[0.2cm]
Conf.	&	$LS$	&	e34	&	e44	&	e52	&	e62	&	NIST \cite{kramida2020nist} 	\\ \hline\addlinespace[0.2cm]
$6s^2$	&	$^1S_0$	&	0 	&	0 	&	0 	&	0 	&	0	\\
$6s6p$	&	$^3P_0$	&	37343 	&	37363 	&	37329 	&	37315 	&	37644.982	\\
	&	$^3P_1$	&	39071 	&	39090 	&	39057 	&	39044 	&	39412.237	\\
	&	$^3P_2$	&	43294 	&	43304 	&	43274 	&	43262 	&	44042.909	\\
	&	$^1P_1$	&	55374 	&	55387 	&	55344 	&	55333 	&	54068.6829	\\
$5d^9 6s^2 6p$	&	$^3P_2$	&	67847 	&	67376 	&	67411 	&	67452 	&	68886.43	\\
$5d^9 6s^2 6p$	&	$^3D_3$	&	69458 	&	68969 	&	69002 	&	69046 	&	70932.2	\\ \hline\hline 
	\end{tabular}}
\end{table*}	

\begin{table*}[htp]
\centering		
\caption{Excitation energies (EE) in cm$^{-1}$ for the ground and lower excited states of Hg atoms calculated using EOM-RCC and different correlation models. Convergence for a larger basis set was considered under the e62-lt100 model, where lt100 means a virtual orbital cutoff energy of 100 a.u..\label{tab:Hg-basis}}{\setlength{\tabcolsep}{10pt}
\begin{tabular}{lllllll}\hline\hline  \addlinespace[0.1cm]
	&		&	 \multicolumn{3}{c}{e62-\texttt{lt100}}					&		\\  \cline{3-5}  \addlinespace[0.2cm]
Conf.	&	$LS$	&	dyall.ae4z	&	dyall.aae4z	&	s-aug-dyall.aae4z	&	NIST \cite{kramida2020nist} 	\\ \hline\addlinespace[0.2cm]
$6s^2$	&	$^1S_0$	&	0 	&	0 	&	0 	&	84184.15	\\
$6s6p$	&	$^3P_0$	&	37479 	&	37473 	&	37454 	&	37644.982	\\
	&	$^3P_1$	&	39277 	&	39268 	&	39249 	&	39412.237	\\
	&	$^3P_2$	&	43816 	&	43797 	&	43777 	&	44042.909	\\
	&	$^1P_1$	&	54823 	&	54754 	&	54712 	&	54068.6829	\\
$6s7s$	&	$^3S_1$	&	64378 	&	62228 	&	62217 	&	62350.325	\\
	&	$^1S_0$	&	67329 	&	64174 	&	64152 	&	63928.12	\\
$5d^9 6s^2 6p$	&	$^3P_2$	&	68343 	&	68226 	&	68302 	&	68886.43	\\
$6s7p$	&	$^3P_0$	&		&	69794 	&	69741 	&	69516.576	\\
	&	$^3P_1$	&		&	69937 	&	69883 	&	69661.803	\\
$5d^9 6s^2 6p$	&	$^3D_3$	&	70175 	&	70168 	&	70266 	&	70932.2	\\ \hline\hline
	\end{tabular}}
\end{table*}

Table~\ref{tab:fs} presents the $F$ factors for Hg and Cn atoms, calculated using both the EOM-RCC and CI+MBPT methods. For Hg, both methods demonstrate consistency in predicting $F$ factors, except for the $6d^9 7s^2 7p\ ^3P_1$ state, where opposite signs are observed between the two methods; and for the $6d^9 7s^2 7p\ ^3D_3$ state, where the CI+MBPT method appears to underestimate the value. For Cn, the agreement between the EOM-RCC and CI+MBPT results is also good, although discrepancies are found for specific states such as $6d^9 7s^2 7p\ ^3P_1$. Compared to Hg, the two methods exhibit overall larger deviations for Cn. To evaluate the accuracy of our calculations, the $F$ factors for three key transitions in the Hg atom are compared in Table~\ref{tab:fsHg}. For the $6s^2\,{}^1S_0 \rightarrow 6s6p\,{}^3P_0$ transition at 265~nm, the EOM-RCC calculation is in consistent with the CI+MBPT result and the multiconfiguration Dirac--Hartree--Fock (MCDHF) result, demonstrating agreement across different high-precision theoretical approaches. The 254~nm transition ($6s^2\,{}^1S_0 \rightarrow 6s6p\,{}^3P_1$) exhibits similar consistency between methods. These values agree with the earlier multiconfiguration Dirac–Fock (MCDF) result~\cite{torbohm1985state} and the more recent MCDHF calculation~\cite{schelfhout2022multiconfiguration}. Experimental measurements using different techniques report a range of values, from muonic X-ray~\cite{hahn1979experimental} to semi-empirical~\cite{ulm1986isotope}, with the present theoretical results falling within this range. For the $6s6p\,{}^3P_2 \rightarrow 6s7s\,{}^3S_1$ transition at 546~nm, both the EOM-RCC and CI+MBPT methods predict positive $F$ values, consistent with the semi-empirical value \cite{ulm1986isotope}. The agreement between these independent theoretical and experimental results  validates the reliability of the methodologies employed in this work.

In Table~\ref{tab:alpha}, the static scalar ($\alpha^S$) and tensor ($\alpha^T$) electric dipole polarizabilities of the ground and excited states in Hg and Cn are presented. The $4\zeta^d$ basis set is not employed due to the high computational cost. Instead, the basis set effect is evaluated by comparing the results obtained with the $3\zeta^s$ and $4\zeta^s$ basis sets. The agreement between these two sets indicates convergence, suggesting that the use of higher-order augmented diffuse functions in the $4\zeta^d$ basis would not lead to improvements at the current level of accuracy. Therefore, the final values are taken from the $4\zeta^s$ calculations. The uncertainty is estimated as the finite basis set size effect and is quantified by the difference between the results obtained with the $3\zeta^s$ and $4\zeta^s$ basis sets. For the ground state of Hg, the EOM-RCC value is found to be consistent with the relativistic coupled-cluster with single, double, and perturbative triple excitations (RCCSD(T)) results reported by Sahoo et al.~\cite{sahoo2018role} and Cheng~\cite{cheng2025relativistic}. For the clock transition $^1S_0 \rightarrow\ ^3P_0$ in Hg, the differential scalar polarizability $\Delta\alpha^S$ is determined to be $20.75(45)\ \mathrm{a.u.}$, which confirms the previously reported value basd on the CI+MBPT method~\cite{hachisu2008trapping}. It is worth noting that Ref.~\cite{hachisu2008trapping} reports a fractional BBR shift of $1.60 \times 10^{-16}$ with an uncertainty of $2 \times 10^{-19}$, assuming a temperature variation of 0.1 K, but without accounting for the uncertainty in the BBR coefficient. This omission leads to an underestimation of the total uncertainty associated with the BBR shift. The BBR shift of the Hg clock transition can be calculated by using,
\[
\delta \omega_{\mathrm{BBR}} = -\frac{1}{2} \left( \frac{2\pi \alpha T}{\nu_0} \right)^2 \Delta\alpha^S,
\]
where $\delta\omega_{\mathrm{BBR}}$ denotes the relative BBR frequency shift of the clock transition, $\alpha$ is the fine-structure constant, $T$ is the temperature, and $\nu_0$ is the absolute frequency of the clock transition in Hg. The value of $\Delta\alpha^S$ calculated in this work, $20.75(45)$ a.u., guarantees only a $3 \times 10^{-18}$ fractional uncertainty under the experimental conditions described in Ref.~\cite{hachisu2008trapping}. This further underscores the importance of accurately determining $\Delta\alpha^S$ and evaluating its associated uncertainty. The present calculation of $\Delta\alpha^S$ for the Hg clock transition supports the development of the Hg optical lattice quantum frequency standard.


The Cn atomic system exhibits distinct relativistic characteristics that influence its polarizability behavior. The ground state $7s^2\ ^1S_0$ displays a reduced polarizability compared to Hg, reflecting the large relativistic contraction of the $7s$ orbitals in superheavy elements. In Table~\ref{tab:alpha}, the $\alpha_S$ and $\alpha_T$ values are reported for many excited states of Hg and Cn, for which no literature values are currently available to the best of our knowledge. Our results for the ground states of Cn and the excited states of Hg are consistent with previously reported data, thereby supporting the reliability of our predicted $\alpha$ values for the excited states of Cn.


\section{Conclusion}

In conclusion, we present a relativistic investigation of the atomic properties of Hg and Cn, employing both the EOM-RCC and the CI+MBPT methods. The work provides reliable values for IP, EEs, $F$ factors and $\alpha$ values for the ground state and several low-lying excited states. For Hg, the calculated results exhibit agreement with available experimental data, thereby supporting the validity of the theoretical approaches and computational strategies adopted. For Cn, the calculations constitute a systematic investigation that includes $F$ factors and $\alpha$ values in excited-state, which to the best of our knowledge have not been previously reported. The analysis of the two theoretical methods confirms their consistency for low-lying states, while exhibit some deviations for some excited states, highlighting the influence of basis set selection and the computational challenges intrinsic to superheavy elements. The results reported here offer a useful theoretical benchmark for future efforts in high-precision spectroscopy, atomic clock development, and fundamental studies involving superheavy element Cn.

\section*{Acknowledgments}
This work is supported by the Innovation Program for Quantum Science and Technology (Grant No. 2021ZD0300901), the National Key Research and Development Program of China (Grant No. 2021YFA1402104), the Cultivation Project for Basic Research and Innovation of Yanshan University (Grant No. 2022LGZD001), and the Space Application System of the China Manned Space Program. J. H. and Z. H. are also supported by the Jiangsu Province Youth Science and Technology Talent Support Project.

\appendix
\section{Selection of Computational Models and Basis Sets
Assessment of Convergence Criteria}	


In Appendix, we provide details for analyzing the influence of various calculation parameters on the EOM-RCC results, including basis sets and electron correlations. Table~\ref{tab:Hg-core} presents calculations employing the dyall.ae2z basis set with a virtual orbital energy cutoff of 10 a.u. (denoted as \texttt{lt10}). Four different electron correlation models are considered: i) the \texttt{e34} model correlates 34 electrons from the $4f$, $5s$, $5p$, $5d$, and $6s$ shells; ii) the \texttt{e44} model correlates 44 electrons from the $4d$, $4f$, $5s$, $5p$, $5d$, and $6s$ shells; iii) the \texttt{e52} model correlates 52 electrons from the $4s$, $4p$, $4d$, $4f$, $5s$, $5p$, $5d$, and $6s$ shells; and iv) the \texttt{e62} model correlates 62 electrons from the $3d$, $4s$, $4p$, $4d$, $4f$, $5s$, $5p$, $5d$, and $6s$ shells. To examine basis set convergence, Table~\ref{tab:Hg-basis} presents the computational results using the e62-\texttt{lt100} model, which correlates 62 electrons with a relaxed virtual orbital cutoff of 100 a.u., thereby providing a more comprehensive treatment of electron correlations. These systematic computations facilitate the selection of the final model and the evaluation of the uncertainties.

\bibliography{apssamp}

\end{document}